\documentclass{elsart}

\usepackage{cite}

\usepackage{dcolumn}

\usepackage{graphicx}

\begin{document}

\begin{frontmatter}

\title{Solution to the Equations of the Moment Expansions}
\author{Paolo Amore\thanksref{PA}}
\address{Facultad de Ciencias, Universidad de Colima, \\
Bernal D\'{\i}az del Castillo 340, Colima, Colima, Mexico}

\thanks[PA]{E--mailpaolo.amore@gmail.com}

\author{Francisco M. Fern\'{a}ndez\thanksref{FMF}}

\address{INIFTA (UNLP, CCT La Plata-CONICET), Divisi\'{o}n Qu\'{i}mica Te\'{o}rica,\\
Blvd. 113 y 64 (S/N), Sucursal 4, Casilla de Correo 16,\\
1900 La Plata, Argentina}

\thanks[FMF]{E--mail: fernande@quimica.unlp.edu.ar (Corresponding author)}

\begin{abstract}
We develop a formula for matching a Taylor series about the origin
and an asymptotic exponential expansion for large values of the
coordinate. We test it on the expansion of the generating
functions for the moments and connected moments of the Hamiltonian
operator. In the former case the formula produces the energies and
overlaps for the Rayleigh--Ritz method in the Krylov space. We
choose the harmonic oscillator and a strongly anharmonic
oscillator as illustrative examples for numerical test. Our
results reveal some features of the connected--moments expansion
that were overlooked in earlier studies and applications of the
approach.

\end{abstract}

\end{frontmatter}

\section{Introduction}

Some time ago Horn and Weinstein\cite{HW84} and Horn et
al\cite{HKW85} proposed the calculation of the ground--state
energy of quantum--mechanical systems by means of the Taylor
expansion of the generating function for the cumulants or
connected moments. The main problem of this approach is the
extrapolation of the $t$--expansion for $t\rightarrow \infty $.
Those authors proposed approximate expressions based on Pad\'{e}
approximants that did not produce encouraging results. For that
reason Cioslowski\cite{C87a} suggested the extrapolation by means
of a series of exponential functions. This and other approaches
were discussed and compared by Stubbins\cite{S88}. Cioslowski's
approach leads to a nonlinear system of equations for the
parameters in the exponential expansion that he solved by means of
a systematic algorithm that avoids the explicit calculation of the
unnecessary variables. The resulting approach has since been known
as the connected--moments expansion or CMX. Later,
Knowles\cite{K87} developed an elegant expression for the CMX
approximants to the energy of the ground state in terms of
matrices built from the connected moments. Since then, the CMX has
been applied to a wide variety of problems and has been
extensively discussed and generalized. A complete list of
references is given elsewhere\cite{AFR10}; here we restrict
ourselves to those papers that are relevant to the present
discussion.

In a recent paper Amore et al\cite{AFR10} analyzed the CMX by means of
simple quantum--mechanical models and conjectured that the parameters in the
exponential expansion proposed by Cioslowski\cite{C87a} may give a clue on
the success of the approach. However, the seminal papers on the CMX\cite
{C87a,K87} as well as all the later applications of the method\cite{AFR10}
(and references therein) were focused on the calculation of the energy
avoiding the explicit calculation of the parameters of the exponential
expansion.

The main purpose of this paper is to provide an explicit solution
to the problem of matching a Taylor series about the origin and an
asymptotic exponential expansion at infinity. We apply it to the
nonlinear CMX equations in order to show the usefulness of the
exponential parameters to predict the success of the approach. We
resort to simple quantum--mechanical models that allow the
calculation of a sufficiently great number of connected moments in
order to test the main equations to any desired order of
approximation. In Sec.~\ref{sec:matching} we develop the main
equations for the general problem of matching the two asymptotic
series. In Sec.~\ref{sec:gen_fun} we discuss the generating
functions for the moments and connected moments and apply the main
equations to them. In Sec.~\ref{sec:examples} we carry out a
numerical test of the general results by means of simple
quantum--mechanical models. Finally, in Sec.~\ref{sec:conclusions}
we discuss the results, draw conclusions and propose further
applications of the main equations.

\section{Matching the expansions}

\label{sec:matching}

Suppose that a function $F(t)$ can be expanded in a formal power
series for small $t$
\begin{equation}
F(t)=\sum_{j=0}^{\infty }\frac{(-t)^{j}}{j!}F_{j}  \label{eq:Taylor}
\end{equation}
and in an exponential expansion for large $t$
\begin{equation}
F(t)=\sum_{j=0}^{\infty }d_{j}e^{-te_{j}}  \label{eq:Exp_exp}
\end{equation}
where ${\mathrm Re}(e_{j})>0$. We can match both expansions at
$t=0$ provided that the series in the right--hand side of the
equations
\begin{equation}
F_{k}=\sum_{j=0}^{\infty }d_{j}e_{j}^{k},\;k=0,1,\ldots
\end{equation}
converge.

We are interested in the case that we do not know the parameters $d_{j}$ and
$e_{j}$ of the exponential expansion. Therefore, we try an ansatz of the
form
\begin{equation}
F^{(N)}(t)=\sum_{j=0}^{2N-1}d_{j}e^{-te_{j}}  \label{eq:F^(N)}
\end{equation}
and match its Taylor expansion about $t=0$ with the actual power
series (\ref {eq:Taylor}). In this way we obtain the following
system of $2N$ nonlinear equations with the $2N$ unknowns $d_{j}$
and $e_{j}$:
\begin{equation}
F_{k}=\sum_{n=0}^{2N-1}d_{n}e_{n}^{k},\;k=0,1,\ldots ,2N-1
\label{eq:nonlin_eqs}
\end{equation}

In order to solve equations (\ref{eq:nonlin_eqs}) we consider the auxiliary
system of $N$ linear equations with $N$ unknowns $c_{i}$%
\begin{equation}
\sum_{i=0}^{N-1}\left( F_{i+j+1}-WF_{i+j}\right) c_{i}=0,\;j=0,1,\ldots ,N-1
\label{eq:secular_gen}
\end{equation}
There are nontrivial solutions only if its determinant vanishes
\begin{equation}
\left| F_{i+j+1}-WF_{i+j}\right| _{i,j=0}^{N-1}=0  \label{eq:secular_det_gen}
\end{equation}
that is to say, if $W$ is one of the $N$ roots $W_{0},W_{1},\ldots ,W_{N-1}$
of the characteristic polynomial
\begin{equation}
\sum_{j=0}^{N}p_{j}W^{j}=0
\end{equation}
where the coefficients $p_{j}$ are nonlinear functions of the $F_{k}$.

If we define
\begin{equation}
\gamma _{j}=\sum_{i=0}^{N-1}F_{i+j}c_{i},\;j=0,1,\ldots ,N-1
\end{equation}
then equations (\ref{eq:secular_gen}) reduce to $\gamma _{j+1}=W\gamma _{j}$%
, $j=0,1,\ldots ,N-1$. It follows from this result that $\gamma
_{j}=W^{j}\gamma _{0}$ and
\begin{equation}
\sum_{j=0}^{N}\gamma
_{j}p_{j}=\sum_{i=0}^{N-1}c_{i}\sum_{j=0}^{N}F_{i+j}p_{j}=\gamma
_{0}\sum_{j=0}^{N}p_{j}W^{j}=0
\end{equation}
We realize that the coefficients $p_{j}$ are given by
\begin{equation}
\sum_{j=0}^{N}F_{i+j}p_{j}=0,\;j=0,1,\ldots ,N-1  \label{eq:F_ijpj}
\end{equation}

Taking into account this equation and Eq.~(\ref{eq:nonlin_eqs}) it is clear
that
\begin{equation}
\sum_{j=0}^{N}F_{i+j}p_{j}=\sum_{n=0}^{N-1}d_{n}e_{n}^{i}%
\sum_{j=0}^{N}p_{j}e_{n}^{j}=0
\end{equation}
In other words, the exponential parameters are the roots of the
secular determinant: $e_{n}=W_{n}$, $n=0,1,\ldots ,N-1$. Once we
have these roots the nonlinear equations (\ref{eq:nonlin_eqs})
become linear equations for the remaining unknowns $d_{n}$. There
are $2N$ such equations but we only need $N$ of them; for
concreteness we arbitrarily choose the first $N$ ones. The
occurrence of multiple roots $e_{n}$ reduces the order $N$ of the
ansatz $F^{(N)}(t)$.

The starting point of present proof Eq.~(\ref{eq:secular_gen}) was motivated
by an earlier paper were Fern\'{a}ndez\cite{F08} proved the equivalence
between the Rayleigh--Ritz variation method in the Krylov space and the
connected--moments polynomial approach\cite{B08}.

\section{Generating functions for the moments and connected moments}

\label{sec:gen_fun}

The generating function for the moments of a Hamiltonian operator $\hat{H}$
with respect to a trial or reference state $\left| \phi \right\rangle $ is
\begin{equation}
Z(t)=\left\langle \phi \right| e^{-t\hat{H}}\left| \phi \right\rangle
\label{eq:Z(t)}
\end{equation}
The coefficients of its Taylor expansion
\begin{equation}
Z(t)=\sum_{j=0}^{\infty }\frac{(-t)^{j}}{j!}\mu _{j}  \label{eq:Z(t)_Taylor}
\end{equation}
give the moments $\;\mu _{j}=\left\langle \phi \right| \hat{H}^{j}\left|
\phi \right\rangle $. If the spectrum of $\hat{H}$ is discrete and its
eigenfunctions form a complete set
\begin{equation}
\hat{H}\left| \psi _{j}\right\rangle =E_{j}\left| \psi _{j}\right\rangle
,\;j=0,1,\ldots
\end{equation}
then
\begin{equation}
Z(t)=\sum_{j=0}^{\infty }\left| \left\langle \phi \right| \left. \psi
_{j}\right\rangle \right| ^{2}e^{-tE_{j}}  \label{eq:Z(t)_exp_exp}
\end{equation}
provided that $\left\langle \psi _{i}\right| \left. \psi _{j}\right\rangle
=\delta _{ij}$. Therefore, we can apply the method developed in the
preceding section with $F_{j}=\mu _{j}$. For concreteness we assume that $%
E_{0}\leq E_{1}\leq E_{2}\leq \ldots $.

In this case equations (\ref{eq:secular_gen}) and (\ref
{eq:secular_det_gen}) are the secular equations and secular
determinant, respectively, for the Rayleigh--Ritz method in the
Krylov space\cite{F08} (and references therein). Therefore, the
roots $W_{j}$ are real and for each of them we have the
approximate solution
\begin{equation}
\left| \varphi _{j}\right\rangle =\sum_{i=0}^{N-1}c_{ij}\left| \phi
_{i}\right\rangle ,\;j=0,1,\ldots ,N-1,\;\left| \phi _{i}\right\rangle =\hat{%
H}^{i}\left| \phi \right\rangle  \label{eq:varphi_j}
\end{equation}
where $\left\langle \varphi _{i}\right| \left. \varphi _{j}\right\rangle
=\delta _{ij}$. Besides, we know that the approximate variational
eigenvalues are upper bounds to the exact ones: $%
W_{j}^{(N)}>W_{j}^{(N+1)}>E_{j}$.

The projection operator
\begin{equation}
\hat{P}_{N}=\sum_{j=0}^{N-1}\left| \varphi _{j}\right\rangle \left\langle
\varphi _{j}\right|  \label{eq:P_N}
\end{equation}
satisfies
\begin{equation}
\hat{P}_{N}\left| \phi _{i}\right\rangle =\left| \phi _{i}\right\rangle
,\;i=0,1,\ldots ,N-1
\end{equation}
For the projected Hamiltonian
\begin{equation}
\hat{H}_{N}=\hat{P}_{N}\hat{H}\hat{P}_{N}
\end{equation}
we have
\begin{equation}
\hat{H}_{N}^{j}\left| \phi \right\rangle =\hat{P}_{N}\hat{H}^{j}\left| \phi
\right\rangle ,\;j=0,1,\ldots ,N
\end{equation}
Therefore
\begin{equation}
\left\langle \phi \right| \hat{H}_{N}^{j}\left| \phi \right\rangle
=\left\langle \phi \right| \hat{H}^{j}\left| \phi \right\rangle
,\;j=0,1,\ldots ,2N-1
\end{equation}
The approximate generating function
\begin{equation}
Z^{(N)}(t)=\left\langle \phi \right| e^{-t\hat{H}_{N}}\left| \phi
\right\rangle  \label{eq:Z^(N)(t)}
\end{equation}
exhibits an exponential expansion
\begin{equation}
Z^{(N)}(t)=\sum_{j=0}^{2N-1}\left| \left\langle \phi \right| \left. \varphi
_{j}\right\rangle \right| ^{2}e^{-tW_{j}}
\end{equation}
and its Taylor series about $t=0$ yields the first $2N-1$ exact
moments
\begin{equation}
Z^{(N)}(t)=\sum_{j=0}^{2N-1}\frac{(-t)^{j}}{j!}\mu _{j}+...
\end{equation}
Therefore, if we apply the method of the preceding section the parameters $%
d_{j}$ and $e_{j}$ of the approximate exponential expansion
(\ref{eq:F^(N)}) should be $d_{j}=$ $\left| \left\langle \phi
\right| \left. \varphi _{j}\right\rangle \right| ^{2}$ and
$e_{j}=$ $W_{j}$ if there is no degeneracy. If $W_{j}$ is
$m$--fold degenerate then the coefficient $d_{j}$ will be the sum
of the corresponding $m$ overlaps $\left| \left\langle \phi
\right| \left. \varphi _{j}\right\rangle \right| ^{2}$. It is
surprising that merely matching an exponential--series ansatz and
a Taylor series may lead to the results of the Rayleigh--Ritz
method.

The function
\begin{equation}
E(t)=-\frac{Z^{\prime }(t)}{Z(t)}  \label{eq:E(t)}
\end{equation}
is monotonically decreasing\cite{HW84} and its Taylor expansion about $t=0$
yields the connected moments $I_{j}$:
\begin{equation}
E(t)=\sum_{j=0}^{\infty }\frac{(-t)^{j}}{j!}I_{j+1}  \label{eq:t_exp}
\end{equation}
that one easily obtains by means of the recurrence relation\cite{HW84}
\begin{eqnarray}
I_{1} &=&\mu _{1}  \nonumber \\
I_{j+1} &=&\mu _{j+1}-\sum_{i=0}^{j-1}\left(
\begin{array}{c}
j \\
i
\end{array}
\right) I_{i+1}\mu _{j-i},\;j=1,2,\ldots  \label{eq:I_j(muj)}
\end{eqnarray}

In order to extrapolate $E(t)$ to $t\rightarrow \infty $
Cioslowski\cite {C87a} proposed the exponential--series ansatz
\begin{equation}
E^{(N)}(t)=A_{0}+\sum_{j=1}^{N}A_{j}e^{-b_{j}t}  \label{eq:E^(N)(t)}
\end{equation}
where the unknown parameters $b_{j}$ are supposed to be real and positive.
Matching this expression with the $t$--expansion (\ref{eq:t_exp}) leads to
the set of equations
\begin{eqnarray}
I_{1} &=&\sum_{n=0}^{N}A_{n}  \nonumber \\
I_{k+1} &=&\sum_{n=1}^{N}A_{n}b_{n}^{k},\;k=1,2,\ldots ,2N
\label{eq:I_j_A_j_b_j}
\end{eqnarray}
Arguing as in the preceding section we conclude that the
exponential parameters $b_{j}$, $j=1,2,\ldots ,N$ are the roots of
the pseudo--secular determinant
\begin{equation}
\left| I_{i+j+1}-bI_{i+j}\right| _{i,j=1}^{N}=0  \label{eq:secular_I_ij}
\end{equation}
Once we have the exponential parameters we solve $N$ of the
remaining linear equations (\ref {eq:I_j_A_j_b_j}) for the
coefficients $A_{j}$ and then we obtain $A_{0}$ from the first
equation:
\begin{equation}
A_{0}=I_{1}-\sum_{n=1}^{N}A_{n}  \label{eq:A0_I1-An}
\end{equation}

In order to test the consistency of the main CMX assumption we can try the
alternative ansatz
\begin{equation}
U^{(N)}(t)=\sum_{j=0}^{N}A_{j}e^{-b_{j}t}  \label{eq:U^(N)}
\end{equation}
and verify that there is a stable root $b_{0}$ that approaches
zero as $N$ increases. The corresponding pseudo--secular
determinant is slightly different from the previous one:
\begin{equation}
\left| I_{i+j+2}-bI_{i+j+1}\right| _{i,j=0}^{N}=0  \label{eq:secular_I_ij_b0}
\end{equation}

In this case we expect difficulties in matching the Taylor and
exponential series because the denominator of $E(t)$ exhibits
zeros in the complex $t$--plane $Z(t)=0$. Amore et al\cite{AFR10}
have already discussed this point by means of simple examples and
here we will show that present mathematical formulas are of
considerable help for that purpose.

In the standard implementation of the CMX one does not calculate
the parameters $b_{j}$ explicitly\cite{C87a}. For example,
Knowles\cite{K87} derived the following explicit expression for
the approximant of order $M$ to the coefficient $A_{0}$:
\begin{equation}
A_{0,M}=I_{1}-\left(
\begin{array}{llll}
I_{2} & I_{3} & \cdots & I_{M+1}
\end{array}
\right) \left(
\begin{array}{llll}
I_{3} & I_{4} & \cdots & I_{M+2} \\
I_{4} & I_{5} & \cdots & I_{M+3} \\
\vdots & \vdots & \ddots & \vdots \\
I_{M+2} & I_{M+3} & \cdots & I_{2M+1}
\end{array}
\right) ^{-1}\left(
\begin{array}{l}
I_{2} \\
I_{3} \\
\vdots \\
I_{M+1}
\end{array}
\right)  \label{eq:A_0,M(Ij)}
\end{equation}
where
\begin{equation}
\lim_{M\rightarrow \infty }A_{0,M}=A_{0}=E_{0}
\end{equation}
provided that the method converges.

If we define the matrices $\mathbf{B}=\left(
B_{ij}=b_{j}^{i}\right) _{i,j=1}^{N}$, $\mathbf{A}=\left(
A_{i}\delta _{ij}\right) _{i,j=1}^{N}$ and $\mathbf{I}=\left(
I_{i+j}\right) _{i,j=1}^{N}$ then we can rewrite equations
(\ref{eq:I_j_A_j_b_j}) with $k=2,3,\ldots ,2N$ as $\mathbf{I}=
\mathbf{BAB}^{t}$. Therefore, if the determinant of the square
matrix in equation (\ref{eq:A_0,M(Ij)}) vanishes then

\begin{itemize}

\item one or more roots $b_{j}$ vanish

\item there are multiple roots ($b_{j}=b_{k}=\cdots $)

\item one or more coefficients $A_{j}$ vanish

\end{itemize}
In any such case the approximation of order $N$ reduces to an
approximation of lesser order.

It is not difficult to prove that
\begin{equation}
S(t)^{2}=\frac{Z(t/2)^{2}}{Z(t)}
\end{equation}
satisfies\cite{C87d}
\begin{equation}
\lim_{t\rightarrow \infty }S(t)^{2}=S_{\infty }^{2}=\left| \left\langle \phi
\right. \left| \psi _{0}\right\rangle \right| ^{2}
\end{equation}
From
\begin{equation}
\frac{d}{dt}\ln S(t)^{2}=E(t)-E(t/2)
\end{equation}
one easily derives an approximation to the overlap in terms of the
parameters of the exponential expansion:
\begin{equation}
\ln S_{N}^{2}=\ln \left| \left\langle \phi \right. \left| \phi \right\rangle
\right| ^{2}-\sum_{j=1}^{N}\frac{A_{j}}{b_{j}}  \label{eq:lnSN^2}
\end{equation}
When $\left\langle \phi \right. \left| \phi \right\rangle =1$ this
expression agrees with the one derived by Cioslowski\cite{C87d} except for
the minus sign that is missing in his Eq.~(21). Cioslowski did not use this
expression directly but an equivalent one in terms of matrices built from
the connected moments. Here we will use it in order to test the formulas
derived above for the exponential parameters. For generality we keep the
term $\ln \left| \left\langle \phi \right. \left| \phi \right\rangle \right|
^{2}$ because in some cases our trial functions will not be normalized to
unity.

\section{Illustrative examples}

\label{sec:examples}

In order to test the equations developed in the preceding section in what
follows we apply them to some simple examples where we can carry out
calculations of sufficiently large order.

We first consider the harmonic oscillator
\begin{equation}
\hat{H}=-\frac{d^{2}}{dx^{2}}+x^{2}  \label{eq:H_HO}
\end{equation}
and the unnormalized trial functions
\begin{eqnarray}
\left\langle x\right| \left. \phi _{g}\right\rangle &=&\exp \left( -\frac{%
2x^{2}}{5}\right)  \nonumber \\
\left\langle x\right| \left. \phi _{e}\right\rangle &=&\left( x^{2}-\frac{1}{%
2}\right) \exp \left( -\frac{2x^{2}}{5}\right)  \label{eq:trial_HO}
\end{eqnarray}
already chosen by Amore et al\cite{AFR10} for their analysis of the
convergence properties of the CMX. Table~\ref{tab:ovHO} shows the exact
overlaps $\left| \left\langle \phi \right| \left. \psi _{j}\right\rangle
\right| ^{2}$, $j=0,2,4,6$, for these two trial functions. We appreciate
that $\left| \phi _{g}\right\rangle $ and $\left| \phi _{e}\right\rangle $
exhibit larger overlaps with the ground and second excited state,
respectively.

Table~\ref{tab:Wd_HO_g} shows the parameters $W_{j}$ and $d_{j}$
for the trial function $\left| \phi _{g}\right\rangle $. The
former converge (from above) towards the eigenvalues of the
harmonic oscillator and the latter towards the exact overlaps
shown in Table~\ref{tab:ovHO} in complete agreement with the
general proof given in the preceding section.

Table\-~\ref{tab:A_b0_HO_g} shows the parameters $A_{j}$ and $b_{j}$, $%
j=0,1,2,3$ for the second CMX ansatz $U^{(N)}(t)$ proposed in the preceding
section. Note that the exponential parameter $b_{0}$ tends to zero as $N$
increases suggesting that the CMX applies successfully to this problem.
Table~\ref{tab:Ab1_HO_g} shows the same parameters but with $b_{0}$ set
equal to $0$ as in the first approach $E^{(N)}(t)$. The results of both
tables approach each other as $N$ increases.

Table~\ref{tab:ov_CMX_HO_g} shows that the approximate overlap $S_{N}^{2}$
given by Eq.~(\ref{eq:lnSN^2}) for the unnormalized trial function $\left|
\phi _{g}\right\rangle $ tends to the corresponding exact result in Table~%
\ref{tab:ovHO}. Cioslowski's approach\cite{C87d} applies successfully to
this example.

The second column in Table~\ref{tab:CMX_HO_A_0M_ge} shows that the CMX
converges rapidly towards the ground state as $N$ increases. This success is
unsurprising in the light of the preceding analysis of the CMX parameters $%
b_{j}$. We obtain the same results from equation (\ref{eq:A0_I1-An}) and the
parameters $A_{j}$ given in Table~\ref{tab:Ab1_HO_g}.

Table~\ref{tab:Wd_HO_e} shows the parameters $W_{j}$ and $d_{j}$
for the trial function $\left| \phi _{e}\right\rangle $. The
former converge (from above) towards the eigenvalues of the
harmonic oscillator and the latter towards the exact overlaps
shown in Table~\ref{tab:ovHO}. Since the overlap of the trial
function with the second excited state is larger than the overlap
with the ground state we expect an anomalous behaviour of both
ans\"{a}tze $E^{(N)}$ and $U^{(N)}$ as discussed by Amore et
al\cite{AFR10}. This is in fact the case and some of the
parameters $b_{j}$ for this trial function are negative or
complex. However, the second ansatz $U^{(N)}$
discussed in the preceding section exhibits a small exponential parameter $%
b_{0}$ that appears to tend to zero as $N$ increases. At the same
time, the corresponding coefficient $A_{0}$ tends to the energy of
the second excited state as $N$ increases as sown in
Table~\ref{tab:A0b0_HO_e}. This behaviour is consistent with the
convergence of the CMX to the second excited state shown in the
third column of Table~\ref{tab:CMX_HO_A_0M_ge} and discussed
earlier by Amore et al\cite{AFR10}. Note that the CMX does not
provide bounds to the energies as the Rayleigh--Ritz method
already does.

As a nontrivial example we choose the simple anharmonic oscillator
\begin{equation}
\hat{H}=-\frac{d^{2}}{dx^{2}}+x^{4}  \label{eq:H_AHO}
\end{equation}
and the unnormalized trial functions
\begin{eqnarray}
\left\langle x\right| \left. \phi _{g}\right\rangle &=&\exp \left( -\frac{%
3x^{2}}{2}\right)  \nonumber \\
\left\langle x\right| \left. \phi _{e}\right\rangle &=&\left( x^{2}-\frac{1}{%
4}\right) \exp \left( -\frac{3x^{2}}{2}\right)  \label{eq:trial_AHO}
\end{eqnarray}
also considered by Amore et al\cite{AFR10}. This oscillator is strongly
anharmonic and enables us to calculate as many terms as desired for all the
approximants discussed above.

Table~\ref{tab:Wd_AHO_g} shows the parameters $W_{j}$ and $d_{j}$, $j=0,1,2$
for the trial function $\left| \phi _{g}\right\rangle $. The former converge
(from above) towards the well known eigenvalues as $N$ increases and the
latter provide the overlaps. Since the overlap with the ground state
dominates we predict that the CMX will converge towards this state\cite
{AFR10}. The second column of Table~\ref{tab:A_0M_AHO_ge} shows the great
rate of convergence of the CMX towards the ground state of the anharmonic
oscillator, already calculated by Amore et al\cite{AFR10}. Once again we
appreciate that the CMX does not provide bounds.

Table~\ref{tab:Ab0_AHO_g} shows the parameters for the ansatz
$U^{(N)}$. The parameter $b_{0}$ tends to zero and $A_{0}$ towards
the energy of the ground state of the anharmonic oscillator as $N$
increases. However, spurious roots $b_{j}$ and values of the
corresponding parameters $A_{j}$ appear when $N\geq 3$. We have
just chosen those that follow the reasonable sequences determined
by the results for smaller values of $N$.

Table~\ref{tab:Ab1_AHO_g} shows the parameters for the ansatz $E^{(N)}$.
Note that the agreement between the parameters of the two ans\"{a}tze for
the anharmonic oscillator is not as good as in the case of the harmonic
oscillator. In this case we also obtain apparently spurious roots $b_{j}$
and coefficients $A_{j}$ for $N>4$. For example, when $N=6$ $b_{2}$ and $%
A_{2}$ are the complex conjugates of $b_{3}$ and $A_{3}$, respectively.
Consequently, the complex parts of $A_{2}$ and $A_{3}$ cancel each other in
equation (\ref{eq:A0_I1-An}) that yields a reasonable approach to the
ground--state energy $A_{0}=1.0603680$. We conclude that the parameters $%
A_{j}$ and $b_{j}$ should not necessarily be real and positive for the CMX
approximants (\ref{eq:A_0,M(Ij)}) to converge neatly towards the
ground--state energy.

Table~\ref{tab:ov_CMX_AHO_g} shows that the overlap between the
ground state of the anharmonic oscillator and $\left| \phi_g
\right\rangle$ calculated by means of Eq.~(\ref{eq:lnSN^2}) agrees
with the result of Table~\ref{tab:Wd_AHO_g}. Once again we realize
that the complex parts of the parameters $A_{j}$ and $b_{j}$
cancel out to produce a reasonable real approximation to the
expected result. The occurrence of complex parameters in the
exponential ansatz is not revealed by the approximants
(\ref{eq:A_0,M(Ij)}) and (\ref{eq:lnSN^2}) based on the connected
moments.

\section{Conclusions}

\label{sec:conclusions}

In this paper we propose a simple formula for matching a Taylor
series about $t=0$ and an asymptotic exponential expansion valid
for large $t$. We applied it to the analysis of the extrapolation
of the $t$--expansions for the generating functions of the moments
and connected moments. Obviously, only $Z(t)$ is suitable for
matching both expansions at origin because this function does not
exhibit singular points. Unfortunately, results coming from it are
not size consistent. On the other hand $E(t)$ exhibits
singularities at the zeroes of $Z(t)$ in the complex $t$--plane
that may hinder the extrapolation (see also Amore et al\cite
{AFR10} for other examples). Our formula enables us to test
whether the main assumptions of the CMX are valid for the
reference state chosen for the study of a given
quantum--mechanical problem. We have analyzed two cases for the
harmonic oscillator and two more for an anharmonic oscillator and
have shown that the CMX equations yield better results for the
former which is not surprising. We have also seen that there may
be a great rate of convergence of the CMX approximants
(\ref{eq:A_0,M(Ij)}) even when the parameters in the ansatz
$E^{(N)}(t)$ are complex. This most interesting feature of the CMX
was not revealed by earlier applications of the approach because
they were based on algorithms that bypass the explicit calculation
of the parameters of the ansatz $E^{(N)}(t)$.

Knowles' equation (\ref{eq:A_0,M(Ij)}) for the energy and
Cioslowski's equation (\ref{eq:lnSN^2}) for the overlap are
remarkable ways of bypassing the explicit calculation of the
unnecessary variables in the nonlinear equation (\ref
{eq:nonlin_eqs}). However, we have shown that it is not difficult
to calculate all those variables explicitly and obtain additional
information on the behaviour of the approach.

Finally, we mention that our formula is not restricted to the
analysis of the generating functions for the moments and connected
moments. In future works we will explore its utility in other
problems of physical interest. Just to mention one example, note
that $Z(it)=\left\langle \psi (0)\right| \exp \left(
-it\hat{H}\right) \left| \psi (0)\right\rangle $ is the projection
of the state at time $t$ $\left| \psi (t)\right\rangle =\exp
\left( -it\hat{H}\right) \left| \psi (0)\right\rangle $ onto the
initial state $\left| \psi (0)\right\rangle $. We easily obtain
$Z(it)$ for the harmonic and anharmonic oscillators from the
results of tables \ref{tab:Wd_HO_g}, \ref{tab:Wd_HO_e} and
\ref{tab:Wd_AHO_g}.

\begin{table}[ht]
\caption{Exact overlaps between the trial functions (\ref{eq:trial_HO}) and
the harmonic--oscillator eigenfunctions}
\label{tab:ovHO}
\begin{center}
\par
\begin{tabular}{rll}
\hline
$j$ & $\left| \left\langle \phi _{g}\right| \left. \psi _{j}\right\rangle
\right| ^{2}$ & $\left| \left\langle \phi _{e}\right| \left. \psi
_{j}\right\rangle \right| ^{2}$ \\ \hline
0 & 1.969393167 & 0.006078373974 \\
2 & 0.01215674794 & 1.515878931 \\
4 & 0.0001125624810 & 0.05586468983 \\
6 & 0.000001158050216 & 0.001291015111 \\ \hline
\end{tabular}
\par
\end{center}
\end{table}

\begin{table}[ht]
\caption{Parameters $W_j$ and $d_j$ for the harmonic oscillator
and the trial function $\left|\phi_g \right . \rangle$
(\ref{eq:trial_HO})} \label{tab:Wd_HO_g}
\begin{center}
\par
\begin{tabular}{D{.}{.}{3}D{.}{.}{11}D{.}{.}{11}D{.}{.}{15}D{.}{.}{15}}
\hline
\multicolumn{1}{c}{$N$} & \multicolumn{1}{c}{$W_0 / d_0$} &
\multicolumn{1}{c}{$W_1 / d_1$} &
\multicolumn{1}{c}{$W_2 / d_2$} & \multicolumn{1}{c}{$W_3 / d_3$} \\ \hline
2 & 1.00000699 & 5.006424635 & 9.368568397 & \multicolumn{1}{c}{---} \\
& 1.969404521 & 0.01216455133 & 0.00009457604481 & \multicolumn{1}{c}{---} \\
3 & 1.0000001 & 5.000187579 & 9.027741984 & 13.67205836 \\
& 1.96939336 & 0.01215747214 & 0.000112048591 & 0.000007675514617 \\
4 & 1.00000000 & 5.000004232 & 9.001287128 & 13.07302878 \\
& 1.969393169 & 0.01215677486 & 0.000112568657 & 0.000001130187743 \\ \hline
\end{tabular}
\par
\end{center}
\end{table}

\begin{table}[ht]
\caption{Parameters $b_j$ and $A_j$ of the ansatz (\ref{eq:U^(N)}) for the
harmonic oscillator and the trial function $\left|\phi_g \right. \rangle$ (%
\ref{eq:trial_HO})}
\label{tab:A_b0_HO_g}
\begin{center}
\par
\begin{tabular}{D{.}{.}{3}D{.}{.}{15}D{.}{.}{15}D{.}{.}{15}D{.}{.}{15}}
\hline
\multicolumn{1}{c}{$N$} & \multicolumn{1}{c}{$b_0 / A_0$} &
\multicolumn{1}{c}{$b_1 / A_1$} & \multicolumn{1}{c}{$b_2 / A_2$} &
\multicolumn{1}{c}{$b_3 / A_3$} \\ \hline
2 & 0.002381248414 & 4.198837892 & \multicolumn{1}{c}{---} & \multicolumn{1}{c}{---} \\
& 1.00145413 & 0.02354586947 & \multicolumn{1}{c}{---} & \multicolumn{1}{c}{---} \\
3 & 0.00004431997162 & 4.010093259 & 8.439885 & \multicolumn{1}{c}{---} \\
& 1.000036479 & 0.02471528396 & 0.0002482361748 & \multicolumn{1}{c}{---} \\
4 & 0.0000007311081954 & 4.000338156 & 8.037169546 & 12.76249194 \\
& 1.000000709 & 0.02469399105 & 0.0003029268491 & 0.000002372560526 \\ \hline
\end{tabular}
\par
\end{center}
\end{table}

\begin{table}[ht]
\caption{Parameters $b_j$ and $A_j$ of the ansatz (\ref{eq:E^(N)(t)}) for
the harmonic oscillator and the trial function $\left|\phi_g \right. \rangle$
(\ref{eq:trial_HO})}
\label{tab:Ab1_HO_g}
\begin{center}
\par
\begin{tabular}{D{.}{.}{3}D{.}{.}{13}D{.}{.}{15}D{.}{.}{15}}
\hline
\multicolumn{1}{c}{$N$} & \multicolumn{1}{c}{$b_1 / A_1$} &
\multicolumn{1}{c}{$b_2 / A_2$} & \multicolumn{1}{c}{$b_3 / A_3$} \\ \hline
2 & 4.003491206 & 8.296508793 & \multicolumn{1}{c}{---} \\
& 0.02472188733 & 0.0002743493113 & \multicolumn{1}{c}{---} \\
3 & 4.000086984 & 8.019485444 & 12.58042783 \\
& 0.02469258182 & 0.0003046174966 & 0.000002754219691 \\
4 & 4.000001796 & 8.000821311 & 12.05918785 \\
& 0.02469138991 & 0.0003048780599 & 0.000003706290872 \\ \hline
\end{tabular}
\par
\end{center}
\end{table}

\begin{table}[ht]
\caption{Overlap for the ground state of the harmonic oscillator from Eq.~(%
\ref{eq:lnSN^2})}
\label{tab:ov_CMX_HO_g}
\begin{center}
\par
\begin{tabular}{ll}
\hline
$N$ & $S_N^2$ \\ \hline
2 & 1.969399291 \\
3 & 1.969393256 \\
4 & 1.969393168 \\ \hline
\end{tabular}
\par
\end{center}
\end{table}

\begin{table}[ht]
\caption{Convergence of the CMX for the harmonic oscillator and the two
trial functions $\left| \phi_g \right. \rangle$ and $\left| \phi_e \right.
\rangle$ (\ref{eq:trial_HO})}
\label{tab:CMX_HO_A_0M_ge}
\begin{center}
\par
\begin{tabular}{lll}
\hline
$N$ & $A_{0,N}(g)$ & $A_{0,N}(e)$ \\ \hline
1 & 1.000304878 & 4.931822888 \\
2 & 1.000003763 & 5.014793896 \\
3 & 1 & 5.002413906 \\
4 & 1 & 5.001402117 \\
5 & 1 & 4.999955757 \\
6 & 1 & 5.002955554 \\
7 & 1 & 5.000013363 \\
8 & 1 & 5.000011300 \\
9 & 1 & 5.000001215 \\
10 & 1 & 4.999999154 \\ \hline
\end{tabular}
\par
\end{center}
\end{table}

\begin{table}[ht]
\caption{Parameters $W_j$ and $d_j$ for the harmonic oscillator and the
trial function $\left|\phi_e \right. \rangle$ (\ref{eq:trial_HO})}
\label{tab:Wd_HO_e}
\begin{center}
\par
\begin{tabular}{D{.}{.}{3}D{.}{.}{13}D{.}{.}{13}D{.}{.}{13}D{.}{.}{13}}
\hline
\multicolumn{1}{c}{$N$} & \multicolumn{1}{c}{$W_0 / d_0$} &
\multicolumn{1}{c}{$W_1 / d_1$} & \multicolumn{1}{c}{$W_2 / d_2$} &
\multicolumn{1}{c}{$W_3 / d_3$} \\ \hline
2 & 2.911817131 & 5.079618783 & 9.870638470 & \multicolumn{1}{c}{---} \\
& 0.04317577925 & 1.498883666 & 0.03707877419 & \multicolumn{1}{c}{---} \\
3 & 1.060922282 & 5.002758941 & 9.097824481 & 14.14651589 \\
& 0.006448836455 & 1.517139667 & 0.05488390863 & 0.0006658074829 \\
4 & 1.001339803 & 5.000101462 & 9.007425114 & 13.18333763 \\
& 0.006087180960 & 1.515949720 & 0.05588279435 & 0.001209845824 \\
5 & 1.000026913 & 5.000003052 & 9.000381181 & 13.01903413 \\
& 0.006078565317 & 1.515881592 & 0.05586900205 & 0.001287657490 \\ \hline
\end{tabular}
\par
\end{center}
\end{table}

\begin{table}[ht]
\caption{Parameters $b_0$ and $A_0$ for the harmonic oscillator with the
trial function $\left| \phi_e \right. \rangle$ (\ref{eq:trial_HO})}
\label{tab:A0b0_HO_e}
\begin{center}
\par
\begin{tabular}{D{.}{.}{3}D{.}{.}{13}D{.}{.}{13}}
\hline
\multicolumn{1}{c}{$N$} & \multicolumn{1}{c}{$b_0$} & \multicolumn{1}{c}{$A_0$} \\
\hline
2 & 0.01634078866 & 5.013227071 \\
3 & -0.002960622766 & 4.999388680 \\
4 & 0.006737033331 & 4.997247173 \\
5 & 0.0003889752190 & 5.000207908 \\ \hline
\end{tabular}
\par
\end{center}
\end{table}

\begin{table}[ht]
\caption{Parameters $W_j$ and $d_j$ for the anharmonic oscillator (\ref
{eq:H_AHO}) and the trial function $\left| \phi_g \right. \rangle$ (\ref
{eq:trial_AHO})}
\label{tab:Wd_AHO_g}
\begin{center}
\par
\begin{tabular}{D{.}{.}{3}D{.}{.}{13}D{.}{.}{13}D{.}{.}{13}}
\hline
\multicolumn{1}{c}{$N$} & \multicolumn{1}{c}{$W_0/d_0$} &
\multicolumn{1}{c}{$W_1/d_1$} & \multicolumn{1}{c}{$W_2/d_2$} \\ \hline
2 & 1.069780255 & 7.871169487 & 20.45762861 \\
& 0.9487351539 & 0.07314504796 & 0.001446505981 \\
3 & 1.061229046 & 7.516944429 & 17.25938517 \\
& 0.9451196068 & 0.07523345172 & 0.002946453665 \\
4 & 1.060427446 & 7.462353629 & 16.44650531 \\
& 0.9447200769 & 0.07517101138 & 0.003352072665 \\
5 & 1.06036628 & 7.456258219 & 16.28617073 \\
& 0.944686457 & 0.07513217688 & 0.003396340115 \\ \hline
\end{tabular}
\par
\end{center}
\end{table}

\begin{table}[ht]
\caption{Convergence of the CMX towards the ground--state $(g)$ and
second--excited state $(e)$ energies of the anharmonic oscillator (\ref
{eq:H_AHO})}
\label{tab:A_0M_AHO_ge}
\begin{center}
\par
\begin{tabular}{rll}
\hline
$M$ & $A_{0}(g)$ & $A_{0}(e)$ \\ \hline
5 & 1.060692159 & 7.439371257 \\
10 & 1.060363186 & 7.456069907 \\
15 & 1.060362073 & 7.450017954 \\
20 & 1.060362093 & 7.451366303 \\
25 & 1.060362090 & 7.455118704 \\
30 & 1.060362090 & 7.454183973 \\
35 & \multicolumn{1}{c}{"} & 7.451642486 \\
40 & \multicolumn{1}{c}{"} & 7.454364274 \\
50 & \multicolumn{1}{c}{"} & 7.454214745 \\
60 & \multicolumn{1}{c}{"} & 7.453864737 \\
70 & \multicolumn{1}{c}{"} & 7.455066766 \\
80 & \multicolumn{1}{c}{"} & 7.455185890 \\
90 & \multicolumn{1}{c}{"} & 7.453941990 \\
100 & \multicolumn{1}{c}{"} & 7.453833053 \\ \hline
exact & 1.060362090 & 7.455697938 \\ \hline
\end{tabular}
\par
\end{center}
\end{table}

\begin{table}[ht]
\caption{Parameters $b_j$ and $A_j$ of the ansatz (\ref{eq:U^(N)}) for the
anharmonic oscillator (\ref{eq:H_AHO}) with the trial function $\left|
\phi_g \right. \rangle$ (\ref{eq:trial_AHO})}
\label{tab:Ab0_AHO_g}
\begin{center}
\par
\begin{tabular}{D{.}{.}{3}D{.}{.}{13}D{.}{.}{13}D{.}{.}{13}}
\hline
\multicolumn{1}{c}{$N$} & \multicolumn{1}{c}{$b_0/A_0$} &
\multicolumn{1}{c}{$b_1/A_1$} & \multicolumn{1}{c}{$b_2/A_2$} \\ \hline
2 & 0.08325285817 & 6.885234502 & 20.86679836 \\
& 1.099547538 & 0.4729461665 & 0.01083962814 \\
3 & 0.001993821505 & 6.359516968 & 17.31475086 \\
& 1.06101123 & 0.4997973255 & 0.02232719076 \\
4 & -0.003775464406 & 6.306016728 & 16.74692001 \\
& 1.057884464 & 0.5006911384 & 0.02434537896 \\ \hline
\end{tabular}
\par
\end{center}
\end{table}

\begin{table}[ht]
\caption{Parameters $b_j$ and $A_j$ of the ansatz (\ref{eq:E^(N)(t)}) for
the anharmonic oscillator (\ref{eq:H_AHO}) with the trial function $\left|
\phi_g \right. \rangle$ (\ref{eq:trial_AHO})}
\label{tab:Ab1_AHO_g}
\begin{center}
\par
\begin{tabular}{D{.}{.}{3}D{.}{.}{11}D{.}{.}{13}D{.}{.}{13}}
\hline
\multicolumn{1}{c}{$N$} & \multicolumn{1}{c}{$b_1/A_1$} &
\multicolumn{1}{c}{$b_2/A_2$} & \multicolumn{1}{c}{$b_3/A_3$} \\ \hline
2 & 6.470844472 & 19.19699588 & - \\
& 0.503507803 & 0.01645848089 & - \\
3 & 6.343642152 & 17.18884282 & 34.42060985 \\
& 0.5002800387 & 0.02285212523 & 0.0002229752367 \\
4 & 6.342827268 & 17.16971044 & 34.25122888 \\
& 0.5002367582 & 0.02290363548 & 0.0002309382854 \\
5 & 5.692034185 & 5.980920104 & 16.54089332 \\
& -0.5783779771 & 1.075044172 & 0.02550181868 \\ \hline
\end{tabular}
\par
\end{center}
\end{table}

\begin{table}[ht]
\caption{Overlap for the ground state of the anharmonic oscillator
from Eq.~(\ref{eq:lnSN^2})} \label{tab:ov_CMX_AHO_g}
\begin{center}
\par
\begin{tabular}{ll}
\hline
$N$ & $S_N^2$ \\ \hline
2 & 0.9459076757 \\
3 & 0.9444614836 \\
4 & 0.9444538767 \\
5 & 0.9449417075 \\
6 & 0.9446880500 \\ \hline
\end{tabular}
\par
\end{center}
\end{table}

\end{document}